\documentclass{article}
\usepackage{spconf,amsmath,graphicx}
\usepackage{multirow, makecell,multirow,booktabs}

\title{GMM-ResNet2: Ensemble of Group ResNet Networks for Synthetic Speech Detection}
\name{Zhenchun Lei, Hui Yan, Changhong Liu, Yong Zhou, Minglei Ma
}
\address{School of Computer and Information Engineering, Jiangxi Normal University, Nanchang, China}
\begin{document}
\ninept
\maketitle
\begin{abstract}
Deep learning models are widely used for speaker recognition and spoofing speech detection. We propose the GMM-ResNet2 for synthesis speech detection. Compared with the previous GMM-ResNet model, GMM-ResNet2 has four improvements. Firstly, the different order GMMs have different capabilities to form smooth approximations to the feature distribution, and multiple GMMs are used to extract multi-scale Log Gaussian Probability features. Secondly, the grouping technique is used to improve the classification accuracy by exposing the group cardinality while reducing both the number of parameters and the training time. The final score is obtained by ensemble of all group classifier outputs using the averaging method. Thirdly, the residual block is improved by including one activation function and one batch normalization layer. Finally, an ensemble-aware loss function is proposed to integrate the independent loss functions of all ensemble members. On the ASVspoof 2019 LA task, the GMM-ResNet2 achieves a minimum t-DCF of 0.0227 and an EER of 0.79\%. 
On the ASVspoof 2021 LA task, the GMM-ResNet2 achieves a minimum t-DCF of 0.2362 and an EER of 2.19\%, and represents a relative reductions of 31.4\% and 76.3\% compared with the LFCC-LCNN baseline.

\end{abstract}
\section{Introduction}
\label{sec:intro}

Speech synthesis and voice conversion (VC) techniques can be used to deceive the automatic speaker verification (ASV) system. The task of synthetic speech detection is to develop an anti-spoofing system that effectively discriminates between spoofed speech and genuine speech.

The deep learning methods based on convolutional neural network (CNN) have shown impressive performance in speech anti-spoofing. The residual convolutional neural network (ResNet) \cite{7780459} are widely used to learn deep speech representation. Chen et al. \cite{Pindrop2021} trained ResNet18 based systems for spoofing detection and achieved very competitive results on the ASVspoof 2021 LA task. ECAPA-TDNN \cite{desplanques20_interspeech} introduces the 1-dimensional Res2Net modules with impactful skip connections and Squeeze-and-Excitation blocks to explicitly model channel interdependencies. RawNet2 \cite{Rawnet22021} is an end-to-end network architecture which directly takes raw speech waveform as inputs and uses six residual blocks in the embedding extractor. AASIST \cite{9747766} uses the RawNet2-based encoder to extract high-level feature maps from raw input waveforms and proposes a variant of the graph attention layer. 

The group convolution has been used as a replacement of standard convolution in constructing multi-branch net architectures. These building blocks are often utilized as templates to build deeper networks for achieving stronger modeling capacity. It is first proposed in AlexNet \cite{AlexKrizhevskySH12} for distributed computing of convolutions in CNN over multiple GPUs. It was shown that the group convolution is very effective on reducing both the number of parameters and the training time of CNN, and could also be used to improve classification accuracy.  Tianyan Zhou et al. \cite{9383531} investigated the effectiveness of ResNeXt for speaker verification, and the second convolutional layer in the ResNeXt block is a multi-branch transformation with different cardinalities. ResNet can also be thought of as a category of two-branch networks where one branch is the identity mapping. In the Transformer \cite{VaswaniSPUJGKP17} model, the multi-head attention allows the model to jointly attend to information from different representation subspaces at different positions, and this is also a grouping technique. 

Deep neural network ensembles have become attractive learning techniques with better generalizability over individual models. Even if every model learned from the same dataset, the advantage gained by averaging the models allows us to obtain robust and stable solution points. There are theoretical and empirical evidence that diversity in error distributions across member models can boost ensemble performance. The score-level ensemble mothod have been used for anti-spoofing. Zhang et al. \cite{zhang22f_interspeech} introduced a general norm-constrained score-level ensemble method that can improve robustness to zero-effort impostors and spoofing attacks by jointly processing the scores extracted from the ASV and CM subsystems. Tomilov et al. \cite{STC2021} proposed a weighted score-level ensemble system which contains LCNN9, ResNet18, and RawNet2. The grouping technique is also considered as an ensemble learning method that implements the idea of training sub-models on feature space subsets. 

In our previous works \cite{lei_icassp22, lei23_interspeech}, the GMM-ResNet model and the grouping technique are proposed for spoofing speech detection. In this paper, we propose GMM-ResNet2, which is based on the  GMM-ResNet and has four improvements.
\begin{itemize}
\item Firstly, the different order GMMs have different capabilities to form smooth approximations to the feature distribution, and multiple GMMs are used to extract multi-scale Log Gaussian Probability features. 
\item Secondly, the grouping technique is used to improve the classification accuracy by exposing the group cardinality while reducing both the number of parameters and the training time. The final score is obtained by ensemble of all group classifier outputs using the averaging method. 
\item Thirdly, the residual block is improved by including one activation function and one batch normalization layer. 
\item Finally, an ensemble-aware loss function is proposed to integrate the independent loss functions of all members.
\end{itemize}
The paper is organized as follows: Section 2 explain the architecture of  GMM-ResNet2 model. The experiments are described in section 3. Finally, the conclusion is given in section 4.

\begin{figure*}
	\centering
	\includegraphics[width=\linewidth]{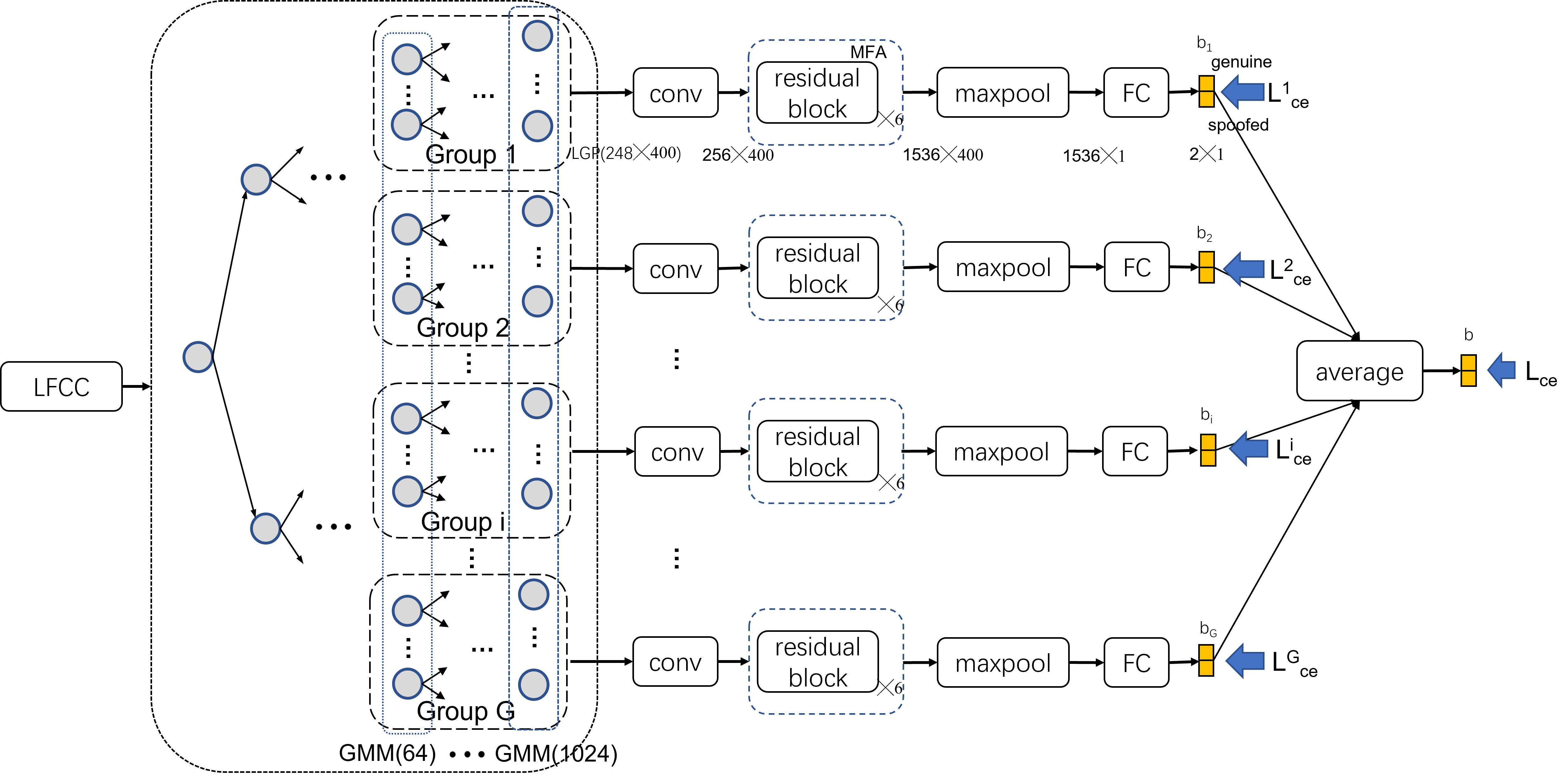}
	\caption{The architecture of GMM-ResNet2.}
	\label{fig:flowchart}
\end{figure*}
\section{GMM-ResNet2}
\label{sec:gmm-resnet2}

The architecture of the proposed model is shown in Figure \ref{fig:flowchart}. The Linear Frequency Cepstrum Coefficients (LFCC) feature is used as the input of the multiple GMMs, and the GMMs extract the Log Gaussian Probability (LGP) features. Then, the LGP features are divided into $G$ groups, and the ResNets followed by a AdaptiveMaxPooling module extract the sub-embeddings respectively. After that, the fully connected layer is used for spoofing speech detection in each group. Finally, the group scores are ensemble using the averaging method.

\subsection{Multiple GMMs } 
In the previous works \cite{lei_icassp22, wen22_interspeech}, the GMM takes raw feature as input and outputs the log probability feature provided by each Gaussian component. For a raw LFCC feature $x$, the element $y_{i}$ of the LGP feature $y$ is defined as:
\begin{equation}
	y_{i} =-\frac{1}{2}{x}'\Sigma_{i}^{-1}x+{x}'\Sigma_{i}^{-1}{\mu_i}
\end{equation}
where $\mu_{i}$ and $\Sigma_{i}$ are the mean vector and covariance matrix of the $i$-th component in GMM. After that, the mean and variance normalization is used.

The LGP feature is based on the GMM, and the LGP size is determined by the order of GMM. The GMM provides an effective way to describe the speech characters, and one of its powerful attributes is the capability to form smooth approximations to arbitrarily shaped densities. However, the capabilities of different order GMMs are also various. So, we concatenate the multi-scale LGP features produced by the different order GMMs to increasing the capacity of LGP features' representation. In our experiments, the number of components in GMMs are 64, 128, 256, 512, and 1024.

\subsection{Grouping and ensemble}

Grouping technique is widely used in deep learning models for its better performance or computational efficiency. In the previous work \cite{lei23_interspeech}, we group the LGP feature to achieve better performance while keeping the model size manageable. The LGP features are based on the Gaussian components in GMM, so we only need to group all Gaussian components. A simple method is random grouping, where each GMM component is assigned to a random group. But this method does not consider the relationship between Gaussian components, and supposes that all components are independent of each other. 

The GMM is usually trained using binary splitting and expectation-maximization (EM). The binary splitting procedure is used to boot up the GMM from a single component to $K$ components. After each split the model is re-estimated several times using the EM algorithm. Since the successor components of the same predecessor component have approximate parameters, we believe that there is a relationship between them. The Gaussian components come from the same previous branch are assigned to the same group. This grouping method increases the ensemble diversity in the context of model averaging.

After grouping, the grouped LGP features are input into 1-d ResNet-based backbones to extract discriminative embeddings respectively. The ResNet module is composed of 6 residual blocks which has 2 convolutional layers and skip residual connection. The max-pooling operation is applied across the temporal dimension. In previous works \cite{desplanques20_interspeech, Rawnet22021, jung22_interspeech,zhang22h_interspeech,9747768}, the low-level feature maps can also contribute towards the accurate  embedding extraction. So, we also apply the multi-scale feature aggregation (MFA) method to the ResNet module for performance improvement.

 Finally, the extracted embedding is input into a fully connection layer classifier for spoofing speech detection in each group. In the ensemble, all group predictions are averaged to generate the final prediction.

\subsection{Improved residual block}
\begin{figure}
	\centering
	\includegraphics[scale=0.5]{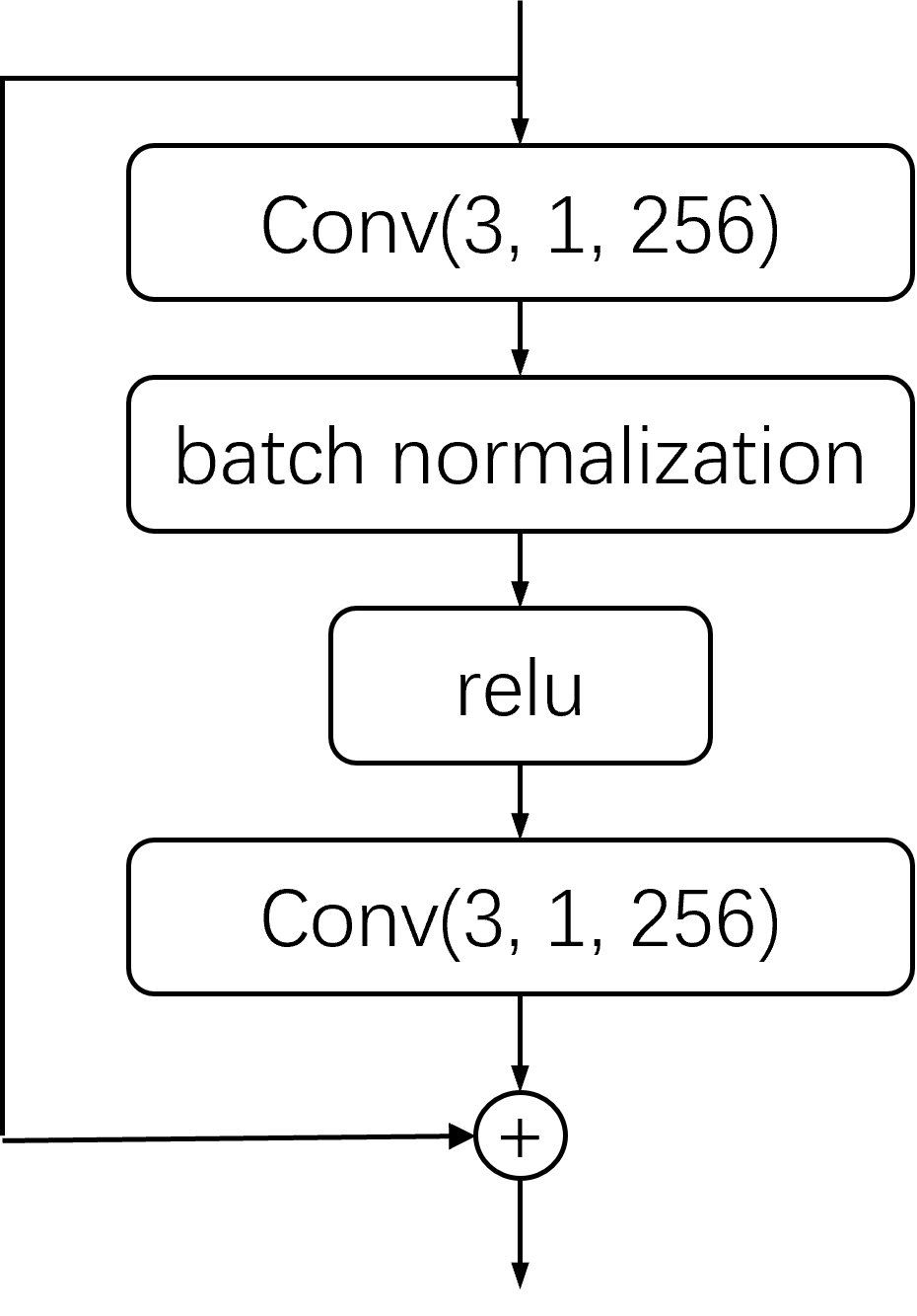}
	\caption{The architecture of residual block. The numbers in parentheses
		refer to kernel size, stride, number of channels.}
	\label{fig:resblock}
\end{figure}

In \cite{9879745}, the authors investigate several layer-wise micro designs. We adopt two design considerations to improve the architecture of residual block: fewer activation functions and fewer normalization layers. One minor distinction between a Transformer and a ResNet block is that Transformers have only one activation function in the MLP block. In comparison, it is common practice to append an activation function to each convolutional layer in residual block. Transformer blocks usually have fewer normalization layers as well.
Figure \ref{fig:resblock} shows the architecture of residual block, which include only one activation function and one batch normalization layer. The activation function and normalization layer following the second convolutional layer have been removed.

\subsection{Ensemble-aware loss function}
we build the ensemble on multiple ResNet-based networks by averaging the output  of each group classifier, as shown in figure \ref{fig:flowchart}. It is natural to explicitly optimize the performance of the corresponding ensemble mean loss during training. We adopt the Cross-Entropy(CE) measurement between the predicted and ground-truth label distributions:
\begin{equation}
	\mathcal{L}_{b}=CrossEntropy(b, y)
\end{equation}
where $b$ is the final output of network, $y$ is the ground-truth label. 

Inspired by \cite{NEURIPS2018_94ef7214}, we add a separate CE loss to each group which simultaneously learn to predict the same ground-truth class label of a training sample. Each group can be considered as an independent classifier in that all of them independently learn high-level semantic representations. Consequently, taking the ensemble of all groups (classifiers) can make a stronger teacher model.

We obtain the overall loss function:
\begin{equation}
	\mathcal{L}=\frac{1}{G+1}(\mathcal{L}_{b}+ \sum_{i=1}^{G}{CrossEntropy(b_i, y)})
\end{equation}
where $b_i$ is the output of $i$-th group classifier, and $G$ is the number of groups.

\section{Experiments}

\subsection{Experimental setup}

 The proposed models are evaluated on the ASVspoof 2019 \cite{ASVspoof2019} logical access (LA) task, ASVspoof 2021 \cite{Yamagishi2021} LA and DeepFake (DF) tasks. According to the evaluation plan, all models are trained using ASVspoof 2019 LA training data, which include 25380 utterances. The evaluation sets of ASVspoof 2021 LA and DF tasks include 181566 and 611829 utterances, respectively. The primary evaluate metric is minimum tandem detection cost function (t-DCF) \cite{tDCF2020}  and the second is the equal error rate (EER). 

 The LFCC is used as the raw acoustic feature in all experiments. The LFCC is extracted following the ASVspoof 2021 baseline configuration \cite{Yamagishi2021}, using a 20 ms window with a 10 ms shift, a 1024-point Fourier transform, and comprising 19 static cepstra plus energy, delta and delta-delta coefficients. The extracted LFCC feature are turned to the fixed length of 400 by truncating or repeating. We train the GMM with 1024 components and 30 EM iterations using the MSR Identity Toolbox  implementation on the ASVspoof 2019 training dataset. The orders of GMMs are 64, 128, 256, 512 and 1024. The size of LGP feature of each utterance is $1984\times400$, which is input into the neural networks. The RawBoost proposed in \cite{9746213} is used to enhance the variation of the training data. New speeches are generated using linear and non-linear (LnL) convolution noise, impulsive signal-dependent (ISD) additive noise, and stationary signal-independent (SSI) additive noise. The number of groups $G$ is set to 8, and the number of channels is 256 in each group residual block.

 The Cross-entropy loss is adopted as the loss criterion, and the Adam optimizer with learning rate of 0.0001 is used during the training phase. The learning rate is adjusted by the ReduceLROnPlateau scheduler. No weight decay is used. The batch size is set to 32, and each model is trained for 100 epochs. Our source codes are publicly available on https://github.com/leizhenchun/gmm-resnet, and all results in this paper are reproducible.
 
 \begin{table}
 	\centering
 	\caption{The number of parameters, multiply-accumulate operations (MACs) of GMM-ResNet and GMM-ResNets2 for a 4-second input.}
 	{
 		\begin{tabular}{ccc}
 			\toprule
 			Model  &  Params(M)  & MACs(G)  \\
 			\midrule
 			GMM-ResNet\cite{lei_icassp22}  & 9.71     & 3.89     \\
 			GMM-ResNet2           & 19.46  & 7.80  \\
 			\bottomrule
 		\end{tabular}
 	}
 	\label{tab:result_model_size}
 \end{table}

Table \ref{tab:result_model_size} shows the number of parameters and multiply-accumulate operations (MACs) of GMM-ResNet and GMM-ResNets2 for a 4-second input.

\subsection{Results on ASVspoof 2019 LA task}

\begin{table}
	\centering
	\caption{Performance of the GMM-ResNet2 on the ASVspoof 2019 LA task in terms of minimum t-DCF and EER (\%). }
	{
		\begin{tabular}{ccc}
			\toprule
			Model  &  min t-DCF  & EER(\%)  \\
			\midrule
			GMM(baseline)\cite{ASVspoof2019}   & 0.2116     & 8.09    \\
			RawGAT-ST\cite{tak21_asvspoof}      & 0.0335     & 1.06      \\
			RawNet2\cite{hansen22_interspeech}        & 0.0289      & 0.99      \\
			ResNet2-FF+PN\cite{10096672}  & 0.027      & 0.94      \\
			ECANet18(SD)\cite{10096837}   & 0.0295     & 0.88      \\
			AASIST\cite{9747766}         & 0.0275      & 0.83      \\
			\midrule
			GMM-ResNet\cite{lei_icassp22}  & 0.0498     & 1.80     \\
			GMM-ResNet2           & \textbf{0.0227}  & \textbf{0.79}  \\
			\bottomrule
		\end{tabular}
	}
	\label{tab:result_19_la}
\end{table}

The ASVspoof 2019 LA scenario contains bona fide speech and spoofed speech data generated by different TTS and VC systems. Table \ref{tab:result_19_la} shows the performance of the GMM-ResNet2 and the most recently known SOTA single systems on the ASVspoof 2019 LA task. The proposed GMM-ResNet2 outperform the baseline system obviously. Compared with the GMM baseline, GMM-Resnet2 can relatively reduce minimum t-DCF and EER by 89.3\% and 90.2\% on the evaluation dataset. Compared with previous GMM-ResNet, GMM-Resnet2 can relatively reduce minimum t-DCF and EER by 54.2\% and 56.1\%. Compared with other SOTA systems, GMM-Resnet2 can achieve better performance. 

\subsection{Results on ASVspoof 2021 LA task}
The ASVspoof 2021 LA task contains speech data generated by different TTS and VC systems with various coding and transmission effects. Table \ref{tab:result_21_la} compares the GMM-ResNet2 with the state-of-the-art methods and four baseline systems (CQCC-GMM, LFCC-GMM, LFCC-LCNN, and RawNet2) provided by the ASVspoof 2021 challenge organizers on LA task. Compared with the LFCC-LCNN baseline, GMM-Resnet2 can relatively reduce minimum t-DCF and EER by 31.4\% and 76.3\% on the evaluation dataset. Compared with GMM-ResNet, GMM-Resnet2 can relatively reduce minimum t-DCF and EER by 4.7\% and 20.9\%. 

\begin{table}
	\centering
	\caption{Performance of the GMM-ResNet2 on the ASVspoof 2021 LA task in terms of minimum t-DCF and EER (\%). }
	{
		\begin{tabular}{ccc}
			\toprule
			Model  & min t-DCF  & EER(\%)  \\
			\midrule
			CQCC-GMM\cite{Yamagishi2021}    & 0.4974	 & 15.62   \\
			LFCC-GMM\cite{Yamagishi2021}    & 0.5758	 & 19.30 \\
			LFCC-LCNN\cite{Yamagishi2021}   & 0.3445     & 9.26     \\
			RawNet2\cite{Yamagishi2021}     & 0.4257     & 9.50     \\
			\midrule
			ECAPA-TDNN \cite{URChen2021}    & 0.3094    & 5.46   \\
			RawNet2+RawBoost \cite{9746213} & 0.3099    & 5.31  \\
			GMM+LCNN \cite{Das2021}         & 0.2672    & 3.62   \\
			ResNet \cite{Pindrop2021}       & 0.2608    & 3.21   \\
            \midrule
			GMM-ResNet                      & 0.2480     & 2.77     \\
			GMM-ResNet2                     & \textbf{0.2362} & \textbf{2.19}     \\
			\bottomrule
		\end{tabular}
	}
	\label{tab:result_21_la}
\end{table}

\subsection{Results on ASVspoof 2021 DF task}
Evaluation data for the ASVspoof 2021 DF task is a collection of utterances processed with different lossy codecs used typically for media storage.
Table \ref{tab:result_21_df} compares the GMM-ResNet2 with the state-of-the-art methods and four baseline systems on the ASVspoof 2021 DF task. 
Compared with the RawNet2 baseline, GMM-Resnet2 can relatively reduce  EER by 26.2\% on the evaluation dataset. Compared with GMM-ResNet, GMM-Resnet2 can relatively reduce EER by 11.0\%. GMM-Resnet2 obtains very competitive results compared with other state-of-the-art systems that are compliant with the ASVspoof training protocol.

\begin{table}
	\centering
	\caption{Performance of the GMM-ResNet2 on the ASVspoof 2021 DF task in terms of EER (\%). }
	{
		\begin{tabular}{cc}
			\toprule
			Model  &  EER(\%)  \\
			\midrule
			CQCC-GMM\cite{Yamagishi2021}    & 25.56     \\
			LFCC-GMM\cite{Yamagishi2021}    & 25.25     \\
			LFCC-LCNN\cite{Yamagishi2021}   & 23.48     \\
			RawNet2\cite{Yamagishi2021}     & 22.38     \\
			\midrule
			ECAPA-TDNN \cite{URChen2021}    & 20.33 \\
			GMM+LCNN \cite{Das2021}         & 18.30 \\
			M-GMM-MobileNet \cite{wen22_interspeech}  &  16.86  \\
			ResNet \cite{Pindrop2021}       & 16.05 \\
			\midrule
			GMM-ResNet         & 18.56     \\
			GMM-ResNet2        & \textbf{16.52}     \\
			\bottomrule
		\end{tabular}
	}
	\label{tab:result_21_df}
\end{table}

\subsection{Ablation experiments}

Ablation experiments were conducted to investigate the impact of each improvement, wherein individual components were removed to explore their contribution to the performance enhancements. The results of the ablation experiments on the ASVspoof 2021 LA and DF tasks are presented in Table \ref{tab:result_ablation}. It is observed that among various strategies employed in GMM-ResNet2, ensemble-aware loss emerges as the most efficient one, as its absence leads to a significant decrease in performance. Furthermore, it is found that the improved residual block slightly enhances performance on the LA task but achieves a remarkable improvement on the DF task.

\begin{table}
	\centering
	\caption{Ablation experiments on the ASVspoof 2021 LA and DF tasks. }
	{
		\begin{tabular}{cccc}
			\toprule
			\multirow{2}*{Model}  & \multicolumn{2}{c}{LA}& \multicolumn{1}{c}{DF} \\
			\cline{2-3}
			&min t-DCF  & EER(\%) & EER(\%)  \\
			\midrule
			GMM-ResNet2                  & \textbf{0.2362}     & \textbf{2.19}  & \textbf{16.52}   \\
			w/o multiple GMMs            & 0.2514     & 2.64  & 18.34   \\
			w/o grouping and ensemble    & 0.2543     & 2.80  & 17.85   \\
			w/o improved residual block  & 0.2375     & 2.21  & 18.13   \\ 
			w/o ensemble-aware loss      & 0.2672     & 3.55  & 19.36   \\
			GMM-ResNet                   & 0.2480     & 2.77  & 18.56   \\
			\bottomrule
		\end{tabular}
	}
	\label{tab:result_ablation}
\end{table}

\section{Conclusions}

 We propose the GMM-ResNet2 architecture for synthetic speech detection in this paper. Compared to the GMM-ResNet, GMM-ResNet2 incorporates four enhancements, including multiple GMMs, grouping and ensemble, improved residual block, and an ensemble-aware loss function. The proposed GMM-ResNet2 shows competitive performance on the ASVSpoof 2019 LA, ASVspoof 2021 LA and DF tasks. In future work, we will further refine the network structure while also investigating advanced grouping methods. We also consider more ensemble models of group networks. And the ensemble-aware loss and the  self-distillation method will also be researched. These models will also be applied to speaker recognition.

\section{Acknowledgements}

This work is supported by National Natural Science Foundation of China (62067004), and by Educational Commission of Jiangxi Province of China (GJJ2200331).



\bibliographystyle{IEEEbib}
\bibliography{refs}

\end{document}